# A REVISED CLASSIFICATION OF ANONYMITY LEVELS

PETER PLEVA

DEPARTMENT OF COMPUTER SCIENCE
UNIVERSITY OF DEBRECEN
H-4010 DEBRECEN, PO BOX 12
`pleva.peter@inf.unideb.hu`

## ABSTRACT

This paper primarily addresses the issue of identifying all possible levels of digital anonymity, thereby allowing electronic services and mechanisms to be categorised. For this purpose, we sophisticate the generic idea of anonymity and, filling a niche in the field, bring the *scope of trust* into the focus of categorisation. One major concern of our work is to propose a novel and universal taxonomy which enables a dynamic, trust-based comparison between systems at an abstract level. On the other hand, our contribution intentionally does not offer an alternative to anonymity metrics, but neither is it concerned with methods of anonymous data retrieval (cf. data-mining techniques). However, for ease of comprehension, it provides a systematic 'application manual' and also presents a lucid overview of the correspondence between the current and related taxonomies. Additionally, as a generalisation of group signatures, we introduce the notion of *group schemes*.

## 1. INTRODUCTION

Following the convention, unless otherwise stated, we adopt *Köhntopp and Pfitzmann*'s terminology [11] in the sequel. For ease of usage, we generally refer to the subject of categorisation as anonymity service (or simply service), irrespective of the underlying protocol[1] or method, the implementing apparatus, or the concrete environment. Still, any kind of application or mechanism can equally well be classified according to the model to be proposed.

Etymologically, the term 'anonymity' originates from the Greek word 'anonymia', simply meaning namelessness. However, that ancient significance has been extended in every sense since the Hellenic era. As an interdisciplinary phenomenon [17] and being mostly used in the context of personal liberty, it now affects our everyday interactions in commerce, in communication, in polling etc. Therefore, it has, in the meantime, taken on many different aspects as well as diverse interpretations. On the other hand, a number of related terms (e.g. identity, pseudonymity, or privacy) have come to exist alongside anonymity throughout the centuries – facilitating refinement of the original concept. Having considered all these facts, it is no wonder that they have led to some *semantic ambiguity* about anonymity. To resolve this inconsistency, we further specify the subject of investigation.

---

[1] By the term 'protocol' we mean concrete cryptographic protocols, meanwhile 'scheme' is used to refer to some protocol scheme.

First, we must determine the type of anonymity service to be concerned with. As pointed out by Díaz et al. [14], there should be made a distinction between data and connection anonymity. While data anonymity is about de-identifying some data, i.e. filtering any personal identifying information out of the data (thereby limiting identity linkage), *connection anonymity* addresses the issue of concealing identities during interaction (incl. the actual data transfer). Since de-identification is predominantly carried out on data sets, it forms the central concern of privacy-preserving data mining (q.v. [12, 13]). From this point on, we strictly confine the discussion to services providing connection anonymity.

Second, it is also crucial to define the type of anonymity that we shall henceforth concentrate on. According to [20], there are three types here to be allowed for. As for environmental anonymity, being determined by external factors (incl. the number and diversity of users) and prior knowledge, its level may vary from situation to situation – even in relation to the same anonymity service. Hence, it cannot be used as a base for our scheme of classification.[2] Meanwhile, *procedural anonymity* of a service deals with the underlying protocol (or method) itself and, as such, depends merely on intrinsic qualities. Consequently, it can be examined through the design of the system and can thus be well adapted to our purpose. Given that the third kind, called content-based anonymity, is about mitigating contextual (i.e. source-related) clues in the transferred data, and thus practically concerns only messaging applications, it can be excluded from our investigation too – for being insignificant for a generic study. All things considered, to construct an abstract categorisation model, it suffices to deal merely with procedural anonymity.

To make it crystal clear, our goal is to present a universally applicable model that, being based solely on scheme-related characteristics, enables dynamic categorisation of services according to the amount of trust[3] that users must place in other participants in order to remain anonymous towards distrusted entities. Dynamicity is to lie in the model's relativity to two investigative parameters: the user role to be observed and the set of distrusted entities. Furthermore, by providing category-level requirement analyses, we shall be able to make the desired model useful to be considered for the design of service schemes, in order for them to ensure a predetermined level of anonymity.

The remaining parts of the paper are organised as follows. Particularising some notable contributions, the succeeding section discusses related work. Afterwards, Section 3 defines anonymity properties, establishes the general principles of our approach, and clarifies terminological questions. Section 4, then, introduces our 5-class categorisation model, describing all corresponding anonymity levels and design requirements in detail. Finally, Section 5 concludes with a concise summary of our findings and exhibits the connection between extant taxonomies and the one to be proposed.

---
[2] As mentioned before, our contribution is not meant for measuring the degree of anonymity (cf. [14,15]).
[3] associating higher levels with potentially less distrust

## 2. RELATED WORK

Although the idea of identifying the levels of anonymity was pioneered by *Flinn and Maurer* as early as 1995, we may ascertain that this very first approach [16] suffices to focus on the context of "networked computer systems", without separating anonymity services from the concrete applications and their settings. Since this global perspective on anonymity enables only situation-dependent examinations to be made, the proposed model cannot be used as a universal means for comparing plain anonymity services. Instead, it was primarily intended to describe all distinct levels of anonymity – involving tangential discussion on implementation issues and requirements. Finally, they concluded by discerning five separate classes.

Following Flinn and Maurer's inaugural investigation, another essential artefact [11], among the few closely-related papers, was published in 2001. In the light of its conventionally-used nomenclature, it is now undisputed that they were *Köhntopp and Pfitzmann* who established the [terminological] building blocks of the field. The most elementary term is pseudonymity, which is introduced to cover the entire spectrum between direct accountability and complete unaccountability (i.e. anonymity in their usage). Contrarily, diverging from Köhntopp and Pfitzmann's approach, our concept of anonymity is identical to that of pseudonymity in their usage (see Section 4). Apart from defining the primitives and standardising the terminology, they also introduced a two-dimensional categorisation of pseudonyms[4]. One dimension classifies pseudonyms according to the context that they are used in. Among these five classes, we make no distinction between person and relationship pseudonyms, nor between role and role-relationship pseudonyms. Henceforward, any persistent pseudonym linkable to a single individual shall simply be referred to as a pseudonym. As for role-related pseudonyms, they fall within the category of group schemes (see Section 3). At the same time, there are three categories[5] along the other dimension, each of which can be directly associated with a particular level of anonymity (see Section 5).

---

[4] Pseudonyms are defined as unique identifiers of subjects (or sets of subjects) and, as such, are suitable to be used to authenticate the holder and their items of interest (e.g. messages, events, actions).
[5] Based on the public's initial knowledge about individual–pseudonym associations, they discerned the following types: public, initially non-public, and initially unlinkable pseudonyms.

Still in 2001, *Seys et al.* made a second attempt [19] to systematically identify the different types of anonymity. After showing that identifiability, linkability, and traceability (see Section 3) are, jointly, not only sensible but also ideal determiners of anonymity, they opted to define categories in terms of these, functionally critical properties. Notwithstanding that five (four practically distinct) types could thereby be recognised, they by themselves, being insufficient to cover all aspects of anonymity, cannot be uniquely associated with specific levels. Therefore, Seys et al. introduced additional properties (conditionality and durability), orthogonal to the defining ones, with the help of which the derived types may be classified according to their strength of anonymity (see Section 5).

The relationships between the above taxonomies are illustrated, in conjunction with a complete enumeration of all potential levels of anonymity (incl. our contribution), in tabular form in Section 5.

## 3. PRELIMINARIES

The aforementioned inconsistency (q.v. Section 1) among definitions of anonymity, which is presumably induced by the term's diffuse meaning and ubiquitous use, perhaps calls for a higher degree of sophistication. Accordingly, resting upon a trust-oriented differentiation, we herein identify each distinct level of anonymity. This classification does not only allow anonymity services to be categorised, but also affords us informal class definitions. Besides, by adopting a pragmatic approach based on narrowing the scope of trust (see Section 4 for details), we are also able to provide some analytical descriptions of the derived classes. Before proceeding to the actual taxonomy, we must, however, clarify some terminological and conceptual questions in order that one can properly apply the model.

**Anonymity Properties**

Similarly to [19], we introduce some analytical properties in order to facilitate characterisation of anonymity levels. We note that certain definitions specified hereunder are borrowed from earlier contributions.

"*Identifiability* is the possibility to know the real identity of some party in the system by means of actual data exchanged in the system." [19]

"*Traceability* is the ability to obtain information about the communicating parties by observing the communication context (e.g. via the IP address)." [21]

"*Unlinkability* of two or more items of interest (IOI) means that within this system, these items are no more and no less related than they are related with respect to the a-priori knowledge." [11]

Although there are situations where it is sensible to distinguish[6] between identifiability and traceability, the two terms can as well be regarded identical in terms of the degree of anonymity; viz. the way by which that identity information is obtained may be indifferent. Therefore, we introduce the generalised concept of *recognisability* as the possibility of 'observing' any personally identifiable information (see next subsection) about individuals, irrespective of the context of observation. Nota bene, we consider personally identifiable information to be observable if it is both accessible and interpretable, most notably not irreversibly obfuscated (e.g. hashed). At the same time, it might still be encoded or encrypted, provided that recoverability requirements are fulfilled, i.e. encoding/encryption mechanisms are identifiable or pre-specified, decryption keys are available etc. Having introduced the above property, we let ourselves concentrate on the fact whether or not these individuals are recognisable, according as there is the possibility of such observation or not. Incidentally, recognisability of individuals implies that they may potentially become accountable for their action. As a remark, we shall still use the more specific terms (identifiability or traceability) to explicitly describe recognisability when occasion requires.

By definition, identifiability and traceability and recognisability can all describe anonymity sharing the same[7] *implicit subject*: the individual. Meanwhile, unlinkability concerns solely IOIs (if not otherwise specified). These, still, do not mean that we cannot talk about traceability (or identifiability) of IOIs or, even, unlinkability between individuals and IOIs – however, in such usages, one should always explicate the desired subject. As regards unidentifiability, untraceability, unrecognisability, and linkability; since their definitions may straightforwardly be deduced from those of the opposite terms, we do not elaborate on them. To exemplify the connections between these terms, let us take some trivial implications: traceable IOIs, which presume observable personally identifiable information, imply linkability between individuals and IOIs, which, in turn, implies individual accountability. Moreover, identifiability, which is a possible source of recognisability, must of necessity proceed from traceable IOIs too. Also trivially, recognisability or, in other words, observability[8] of PII entails linkability.

Notwithstanding that most people would like to be unrecognisable in the electronic world, there is a strong *conflict of interests* over digital unrecognisability. On the one hand, all individuals should be enabled to hide their identities, whilst fraudsters should, on the other hand, be prevented from dodging responsibility for their actions. This conflict between privacy concerns and legal[9] accountability can, however, be reconciled (see limited anonymity). In contrast to unrecognisability, unlinkability is not crucially required to ensure anonymity. Still, it might increase the sense of security of the user, especially when they are subject to recognisability. Nevertheless, linkability, per se, does not directly affect privacy concerns. We observe that there are services which, owing to their intrinsic characteristics, essentially preclude unlinkability (see Subsection 'Pseudonymity').

---

[6] "With respect to identifiability, identity information is acquired through the actual data communicated in the system; traceability focuses on the context of the communication to get this information." [21]

[7] Literally speaking, it is apparently the communication (not the individual) which can be traceable; however the property of traceability, analogously to identifiability, tells us about the recognisability of *the individual*.

[8] To avoid inconsistency with [11], if possible, we shall henceforth shun the use of the term 'observability'.

Given that the degree to which anonymity properties are preserved may by design be limited, according as they can or cannot be violated eligibly (see Subsection 'Scope of Trust'), we should somehow be able to indicate such *conditionality*. Appropriately, when individuals may under no circumstances become recognised, we regard them as unconditionally unrecognisable, and say that unrecognisability is completely preserved.[10] Otherwise, if unrecognisability can eligibly be violated (i.e. it is conditional), individuals are said to be conditionally unrecognisable. Finally, in the case of unauthorised recognisability, we regard unrecognisability as void. We can likewise scrutinise conditionality in connection with providing recognisability, (un)traceability, (un)identifiability, (un)accountability, (un)linkability, or (un)observability of PII. This qualification of security properties allows for a more sophisticated analysis of the levels of anonymity.

**Remark.** Conditional unrecognisability can indicate that some designated participant (q.v. Subsection 'Scope of Trust'), being allowed to trace or link IOIs, may violate complete unrecognisability. On the other hand, it can either imply that identities may be made subject to recognition unintentionally, i.e. when it is precipitated by a fraudulent act. Equally, the same applies to conditional unlinkability. (see limited anonymity)

**Personally Identifiable Information**

Another term that should be brought into central focus is personally identifiable information*, PII* for short. For the purpose of the study, we define PII as information sufficient to uniquely identify, or trace to, a specific individual or, alternatively, a piece of such information. In consequence, PII is simply meant to ensure identifiability of individuals. Examples of PII include[11], but are not limited to, names (such as a full name, a mother's maiden name, or an alias), personal identification numbers (such as a national identification number, a passport number, a driver's license number, a taxpayer identification number, a patient identification number, a financial account number, or credit card number), address information (such as a street or email address), asset information (such as an IP or MAC address), telephone numbers, personal characteristics (such as a facial image or a biometric

---

[9] In some cases, it might suffice to provide merely informal accountability.
[10] The adjectives 'complete' and 'unconditional' are interchangeably used, and so are their derivatives.
[11] according to the recommendations of the U.S. National Institute of Standards and Technology [22]

identifier), information identifying a private property (such as a vehicle registration number or a title number), and any information about an individual that is linkable to one of the above (incl. date of birth, place of birth, and geographical indicators).

In the following of this subsection, we introduce a purely pragmatic classification of PII types. To this, let us concentrate on *resolvability*, i.e. the difficulty of tracing PII to the respective individuals in terms of the accessibility of the data set containing the given PII associations, the so-called *PII records*. Since services applying distinct types of PII may call for distinct degrees of user confidence (see Section 4), we find it reasonable to differentiate between directly resolvable (e.g. most landline telephone numbers), indirectly resolvable (e.g. bank account number), and unresolvable (e.g. DNA fingerprints) PII. For ease of usage, the former two types are respectively referred to as direct and indirect PII. While direct PII requires no external assistance to be traced to the respective individual, in the case of indirect PII, identification may only be carried out with the involvement of a third party having access to the given PII record. By contrast, ensuring against personal recognisability, unresolvable PII must not be traceable to anyone – with the possible exception of the individual themself. In addition, PII may also classify legal accountability as direct or indirect, according as the authorities do or do not need a third party to bring fraudsters to account. One can easily ascertain that, unless managed by some authority, indirect PII straightforwardly entails indirect accountability. Furthermore, we may equally establish a similar correlation in the direct case, viz. public disclosure of direct PII likewise implies direct accountability. Nevertheless, the use of direct (or indirect) PII, per se, does not definitely indicate direct (or indirect) accountability. On the other hand, applying solely unresolvable PII implies unaccountability with no proviso.

**Remarks.** In effect, any PII can be made directly available to the public by listing the corresponding PII records in a public directory. Hence, the above characterisation enables only temporal classification of PII data, viz. any initially indirect (or unresolvable) PII can become directly resolvable (or indirectly resolvable) over time. Although these ruminations are, apparently, highly dependent on what we regard as public (see Subsection 'Scope of Trust'), some types are, by nature, readily available (or initially unavailable) – irrespective of the concrete setting.

**Pseudonymity**

As distinct from [11], where the whole continuum between direct accountability and complete unaccountability is called pseudonymity, we introduce a more specific usage thereof. Be they in a peer-to-peer or in a client-server environment, services (esp. in user-adaptive systems [18]) often require the *use of pseudonyms* to be able to maintain, by observing and personalising user interactions, long-term relationships between entities. Notwithstanding that (persistent) pseudonyms may therefore be advantageous for user modelling, they impede realisation of unlinkability. One might, however, consider introducing user-determined (i.e. initially unlinkable) pseudonyms as means of providing personal unrecognisability and thus unaccountability – in addition to linkability. Note that each pseudonym can be correlated with either type of PII: public pseudonyms class as direct PII, whereas initially non-public and initially unlinkable pseudonyms may respectively be regarded as indirect and unresolvable PII. Furthermore, any PII that is per se appropriate for uniquely indentifying an individual (within a given anonymity set) may potentially be used as, or be translated into (e.g. via encoding), a pseudonym.

Although the concept of PII subsumes pseudonymity, they each represent different aspects of identification. While a pseudonym is one single identifier (i.e. a string of characters) associated to an individual, PII may refer to any person-specific knowledge, fact, or data.[12] Therefore, bringing PII into the focus of examination allows for a more abstract approach to scrutinising identity leakage, most notably it lets us merge identifiability and traceability (q.v. Subsection 'Anonymity Properties'). Additionally, a public or initially non-public pseudonym should of necessity be associated with PII representing the holder, in order for accountability to be ensured.[13] Such associated PII must thus be sufficient to uniquely identify a single individual.

**Remark.** As an ancillary to [11], we observe that not every type of initially unlinkable pseudonym is suitable to perfectly preclude accountability; viz. personal characteristics, being inalienable and tamper-proof, may support allegations. Contrarily, other types of initially unlinkable pseudonyms, come they from asset-specific identifiers (e.g. MAC addresses) or be they actually defined by the user themself (e.g. non-public email addresses), do indeed serve repudiation.

**Investigative Principles**

Prior to each anonymity analysis, we, initially, need to decide from which party's viewpoint we wish to investigate the concrete service. Let us regard to this party as *the observee*. By 'party', we mean to refer to, nota bene, a general representative of a participant role. Obviously, the anonymity set is thus comprised of all potential members of the same role.

**Remark.** By contrast with [19], we do not examine anonymity towards some particular party. Rather, each examination must be carried out towards all distrusted entities (see Subsection 'Scope of Trust').

As to the categorisation, the basic principles to be applied are as follows. If some direct PII of the observee is temporarily or permanently observable, then the service under scrutiny provides no anonymity for them and thus comes under the lowest-level category (see void anonymity). Generally put, if it exhibits characteristics that would imply a lower level of anonymity as well as such that would indicate a higher one, then it clearly falls into the lower class. More simply put, services should be placed in the lowest class of those under consideration, i.e. whose definitions are satisfied. We refer to the preceding principle as the *lowest level principle*.

---

[12] Still, personal identification numbers provide tangible instances of the coincidence between the two notions.
[13] Pseudonym–individual associations can be stored as part of the PII records.

As mentioned above, we do not distinguish between temporary and permanent leakage of PII, evading thereby the question of durability [19]. At the same time, investigations should always be confined to *ordinary circumstances*, i.e. neglecting situations where anonymity ceases by virtue of inappropriate behaviour (see limited anonymity).

**Group Schemes**

By using the term 'party' before, we do not constrain the examination to participant roles: groups, and thus group schemes, can also be studied from the aspect of anonymity. We call a mechanism a *group scheme* if it 1) operates on groups and 2) enables group authentication (cf. membership authentication [9]), i.e. suffices to verify membership (e.g. via credentials) to allow a group action to be performed and 3) enables authenticated members to take measures on behalf of the group and 4) precludes unconditional recognisability as well as unconditional linkability. The introduction of the notion of group schemes is motivated by the numerous alternative derivatives of Chaum's credential scheme [7,8], whose most prominent members are group signatures (originally proposed by Chaum and van Heyst [3]). Owing to the above properties, such group-based schemes, clearly, provide group members with some kind of anonymity: let us call it *group anonymity*. They may also involve a dedicated, trusted member, called group manager; who is responsible for adding (and removing) individuals to (and from) the group and has the ability to reveal identities in disputed cases. We propose drawing distinction between 'plain' and group anonymity, for the simple reason that in the latter case one should exercise more caution in determining the level of anonymity – due to a few extra factors which, as discussed later, need to be taken into account. On the other hand, since group authentication guarantees unrecognisability and thus unlinkability by definition, group schemes obviate low-level[14] anonymity. Accordingly, both untraceability and unlinkability are, inter alia, essential prerequisites for group signatures (as defined by Chaum and Heyst).

**Remarks.** There may be situations where it is sensible, most notably for privacy-enhancing purposes, to incorporate a group scheme into a service. In such cases, we can determine the level of anonymity by investigating solely the embedded mechanism.[15] The easiest (though primitive) way to provide group anonymity is to use role-related pseudonyms (q.v. Section 2) in conjunction with passwords.

---

[14] Let us regard void and apparent anonymity as low, whilst limited and unconditional anonymity as high levels.
[15] If so, the findings shall obviously relate to the viewpoint of that specific group only.

## Scope of Trust

Another crucial issue, which is of prime focus in this work, is the question of publicity, i.e. the extent of exposure of individuals to recognisability. Notwithstanding that enabling people to be truly anonymous should entail preventing any type of PII from being observable to anyone (the observee apart), we cannot have such control over directly resolvable, nor over unresolvable PII – as they are unconditionally observable or, respectively, unobservable. In consequence, we shall henceforth concentrate predominantly on indirectly resolvable PII. As known, it is, for various reasons, not uncommon for an anonymity service to involve trusted third parties (TTP), who are deliberately let in on certain privacy-sensitive information: e.g. TTPs may be designated as group managers in group schemes. Such trusted entities can not only observe PII, but may also be given the exclusive privilege of managing PII records, all without completely compromising anonymity (see revocable anonymity). In addition, there are a number of occasions when it is even inevitable to expose PII. Namely, specific, commonly-used PII types (e.g. personal identification numbers) can equally well be registered and maintained by distrusted third parties (DTP), who may then inherently reveal identities and, thereby, break true anonymity (see apparent anonymity).

In either case, be they trusted or distrusted, third parties should be regarded as loosely-coupled stakeholders of the system (see Figure 1). In addition, direct stakeholders (i.e. participants) may also be classified as trusted or distrusted, according as we do or do not desire to preserve unrecognisability towards them. Therefore, after deciding on the observee, one also needs to determine the set of trusted entities, the so-called *trusted set*. Beside the TTPs, we treat, by default, solely the observee[16] as trusted, all other entities (incl. the public and DTPs) are consequently regarded as distrusted. Nonetheless, trusted sets can be extended to include further trustworthy participants – whilst others should be attached to the distrusted group. Afterwards, we can begin the investigation by discovering which entities the observee is actually forced to be subject to[17] if wanting to remain unrecognised. Let us refer to the result of the discovery as the *scope of trust*. We note that, as an extreme case, the observee may need to rely on themself, in the sense that they must act honestly so as not to become accountable (see forfeitable anonymity). Once some PII is observable from outside the scope of the system, we consider it *publicly known*.

Given that trusted stakeholders may by design be allowed to observe PII, unrecognisability and unlinkability should first be verified against distrusted entities (who may de facto violate them). If this verification fails, we regard the given property as void; otherwise, it should be reverified against the trusted set. We can thus have conditional or unconditional unrecognisability (or unlinkability) according to the outcome of the reverification. As, in our approach, the level of anonymity provided by a service is highly relative to the classification of entities, any modification in the trusted (or in the distrusted) set may potentially affect the result of the examination.

---

[16] taken as an individual at this time
[17] due to their capability to make the observee accountable

**Remarks.** If one would demand a generic constraint on publicity, it should be formulated as follows: preserving true anonymity generally[18] requires that resolvable PII be unobservable to any distrusted entity. In other words, the *publicity constraint* dictates that the scope of trust should be a subset of the trusted set. To refer to a general member of the trusted (or the distrusted) set, we use the term 'entity'.

In summary of the preliminaries, there are two essential parameters which need to be determined preparatory to investigating an anonymity service: the observee and the trusted set. The investigation itself, in turn, consists in discovering the scope of trust and in contrasting it with the trusted set, as discussed in Section 4. In accordance with what was stated above, Figure 1 illustrates how entities should be regarded in the course of the investigation.

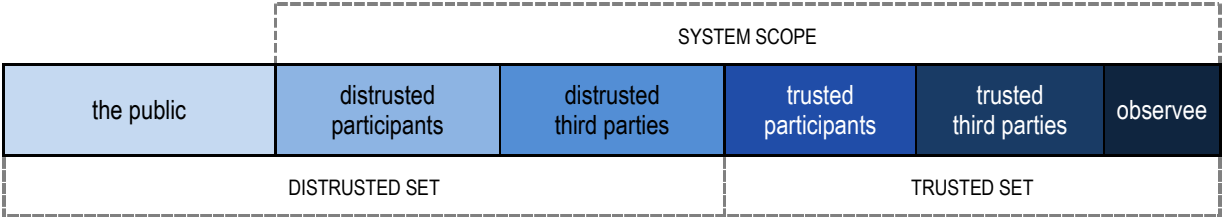

*Figure 1 – A comprehensive classification of entities*

## 4. LEVELS OF ANONYMITY

In the current section, we look at the classification of levels of anonymity induced by the scope of trust. To identify these levels, we concentrate on the recognisability of individuals, or more specifically on the extent to which services expose PII to observation. Clearly, in respect of a concrete investigation, one should confine oneself to examining solely observee-related PII. Let us keep in mind that accessible PII may not necessarily mean readily interpretable plaintext data: it can equally be encrypted or encoded on occasion. Furthermore, as observability of accessible PII is relative to whom it is interpretable for, one should always name this related party (to which the observee's PII is observable) in conjunction – unless it is implicit or indifferent.

### Level 0 (void anonymity – VA)

At one extreme of the scale, people are completely deprived of the possibility of obscuring their identity and may become directly accountable for their actions. This is the case when, violating the publicity constraint, some direct PII data are publicly observable. Accordingly, if a service, for example, exposes public pseudonyms (e.g. unencrypted public keys in a PKI[19]), it cannot be recognised as anonymous – at least not from the given viewpoint. As an implication of unconditional recognisability, void anonymity ensures (unconditional) linkability. Furthermore, the scope of trust involves at least one entity from outside the scope of the system.

---

[18] with one exceptional situation (see revocable anonymity)

**Remark.** In the case of group anonymity a similar situation could trivially arise if the group were limited to either one or two members. However, such contextual matters are dealt with by environmental anonymity, and thus fall out of the scope of this paper. Normally, we cannot have group anonymity at this level.

## Level 1 (apparent anonymity – AA)

A slightly more advanced way of providing anonymity may be the application of indirect PII, which makes calling individuals to account somewhat more complicated. Let us assume that, still in violation of the publicity constraint, these indirect PII data are observable to distrusted entities (e.g. to the public). Then identification can still be carried out, be that a bit more difficult, with the help of some third-party stakeholder[20]. Apparent anonymity presumes that at least one of such entities belongs to the distrusted set – otherwise, one should upgrade to revocable anonymity. The use of initially non-public pseudonyms (e.g. plain credit card numbers) provides an obvious example of this scenario, as long as the above visibility conditions are satisfied. As an alternative scenario of achieving apparent anonymity, we may even introduce direct PII data, with the proviso that they should be publicly unobservable, but be observable to all trusted entities and some distrusted participant.[21] Once, apart the public, there is no distrusted entity, one should proceed to limited anonymity. For instances, we may take encrypted public pseudonyms. Similarly to the preceding level, the requirements for recognisability, and thus for linkability, are unconditionally met. Notwithstanding that the scope of trust again contains some distrusted entities, it excludes the public.

**Remark.** Although group schemes do usually apply a dedicated trusted participant (namely the group manager), that must of necessity be trusted. Moreover, they by definition preclude unconditional recognisability. Consequently, group anonymity even surpasses the level of apparent anonymity.

## Level 2-3 (limited anonymity)

Having taken into consideration the similarity between the concept of extending the trusted set and that of forfeiting the ability to be anonymous, the subsequent two classes are to be jointly discussed.

---

[19] public-key infrastructure
[20] This may be a usual participant as well as an actual TTP or DTP.
[21] It is easy to devise schemes whereby but insiders can, by nature of their position, observe confidential data.

Their analogy straightforwardly arises from the concepts' common, trust-based nature, viz. in the former case they are the users who must place their trust in TTPs, while the latter rests upon the assumption that each individual inclines to obey not only the law, but also corresponding rules and/or policies. In any regard, both levels guarantee anonymity by a relatively narrow, limited scope of trust.

**- Level 2 (revocable anonymity – RA)**

As mentioned previously, PII may be known to dedicated trusted entities without abandoning high-level anonymity. This means that, in spite of the few incidental adverse consequences, it can still be sensible to include trustworthy entities in the trusted set. The rationale for making such a compromise is that we can thus designate TTPs, or other trusted participants, to manage PII records so that any IOI[22] can of necessity be traced back. On the downside, by involving such a so-called *identity manager*, anonymity, or more precisely unrecognisability, can only be conditionally preserved. Another drawback is that individuals must thus not only content themselves with limited anonymity, but also face the threat of identity abuse. Then, beyond the trust that must be put in the identity manager themself, they [the user] also need to rely on the security mechanism deployed by the identity manager. As discussed at apparent anonymity, unhidden indirect PII may promote revocable anonymity, provided that it is observable to distrusted entities, but no identification can be performed without involving an identity manager. Therefore, if satisfying these latter conditions, initially non-public pseudonyms can again exemplify the level. Another scenario, which can be assigned to the same scope of trust, is that we expose direct PII to trusted participants in such a way that it should be unobservable to distrusted entities. Once, apart the observee, there is no trusted entity, one should further proceed to forfeitable anonymity. As regards the appellation, 'revocable' flows from the following consideration. It is not difficult to conceive that, on valid grounds, an identity manager becomes inclined to reveal an identity to a third person, and thereby revokes a user's anonymity. Such a revocation process might be triggered, for instance, 1) in fraud cases when the authorities appeal for an adversary's identity or 2) by disobedience, once some predefined rules and/or policies are violated or 3) by expiry of entitlement to be anonymous. Moreover, anonymity may also be revoked if an individual themself has an interest in revoking their identity.[23] Occasionally, as suggested by the previous arguments, it can be not merely sensible, but also substantive to ensure revocability, and thus only conditional unrecognisability (unidentifiability or untraceability). All things considered, it must be beyond dispute that, at this level, identities are entirely subject to identity managers, thereby being exposed to revocation. Besides, our denomination can be further justified by simplicity and prevalence reasons. Despite that exposing persistent PII may support linkability, we can limit the possibility of associating IOIs by using mutable PII (e.g. dynamic IP addresses) or by reversible obfuscation (after adding some random salt). In addition, applying unconcealed, initially non-public, transaction pseudonyms [11] can also make it a privilege of the identity manager to link IOIs – the latter can by no means be deprived of that capability. The scope of trust, of necessity, includes the observee as well as some further trusted entity (e.g. the identity manager) and comprises, at most, all the trusted entities.

---

[22] e.g. confidential messages or any legal/illegal action taken by a user/adversary

[23] Esp. when they want to be acquitted of a charge or are to be given a reward (e.g. a good mark).

**Remark.** Note that, in group schemes, it is the group manager who may appropriately fill the role of the identity manager. Nonetheless, at this point, there is some difference between group-based and plain anonymity in connection with linkability, viz. it is by definition precluded by the former type. In default of a group manager, one should again consider upgrading to a higher level. Otherwise, this is the lowest level that a group scheme may afford.

**- Level 3 (forfeitable anonymity – FA)**

By forfeitable anonymity, individuals remain anonymous as long as policies, rules, and – not least – the law are followed. Upon violating any of them, the person incriminated forfeits their anonymity. The reasoning behind the idea of forfeiture can be backed by various realistic situations. Forfeitability may be beneficial for example 1) in e-payment protocols to prevent double-spending or 2) in turnstile systems to restrict someone's movements to a definite area (e.g. within the workplace). Allowing anonymity to be subject to forfeiture may obviate inappropriate user behaviour by deterring individuals from committing fraud or from being disobedient. The distinction drawn between revocable[24] and forfeitable[25] anonymity comes from the corresponding term definitions [at footnotes]. Accordingly, revocability requires a participant [identity manager] who is responsible for revocation; whereas forfeitability suffices to require an event-driven forfeiture mechanism[26], as a result of which (direct or indirect) user PII gets observable, and adversaries can become known. Such a typical triggering event would be, e.g., an attempt at double-spending an e-cash coin. Given that, normally, no resolvable PII data are observable, unidentifiability is conditional solely on obedience. The fact, however, remains that IOIs must be able to be traced back after forfeiture; which, in turn, implies that some (interpretable) resolvable PII should be attached to them.[27] Since third parties can legitimately be obliged to reveal identities in fraud cases, indirect PII may as well come into consideration as direct one. In consequence, public and initially non-public pseudonyms, though not specifically intended for that, might equally well be introduced for identification purposes (provided that they are inaccessible and/or uninterpretable until fraud occurs). Moreover, initially unlinkable pseudonyms may also be applied (in addition or solo) to guarantee unconditional linkability. For such purpose, unresolvable PII might thus be introduced and even be exposed to observation.[28] At the same time, as no resolvable PII can be obtained by observing the communication context, traceability cannot come into consideration. The scope of trust, at most, contains the observee.

---

[24] revoke — to annul by recalling or taking back (Merriam-Webster)
[25] forfeit — to lose or lose the right to especially by some error, offense, or crime (ibid.)
[26] Accountability can principally based on a *cut-and-choose* technique or a *zero-knowledge proof* (introduced in [4,5] respectively), as demonstrated by L. Law et al in [6].
[27] Introducing unresolvable PII would, here, be unreasonable.

**Remark.** In accordance with the preceding class definition [revocable anonymity] as well as with the lowest level principle, existence of identity managers precludes forfeitable anonymity. Not to mention that involving such a trusted entity would even make any forfeiture strategy useless (as identities could simply be revoked).

Be it revocable or forfeitable, limited anonymity can not only prevent abuse, but also assure beneficent individuals (and even the authorities) that adversaries can potentially be called to account, whilst personal rights are respected. All aspects considered, limited anonymity seems to be the best compromise between privacy concerns and accountability.

**Remark.** In need of a universal expression for the process of discontinuing anonymity, one may conventionally use *de-anonymisation* – regardless of the concrete level of limited anonymity. It must also be noted that elsewhere in the literature 'revocation' and derived terms are predominantly used in a broader sense, namely as a synonym for de-anonymisation.

### Level 4 (unconditional anonymity – UA)

At the other extreme of the anonymity spectrum (opposite to direct accountability), having made sure that none of the previous class definitions is satisfied, one may recognise a service as unconditionally anonymous. Despite the fact that, being the most advanced way of protecting privacy, unconditional anonymity minimises information leakage, it does not completely exclude the possibility of using PII. Namely, unresolvable PII may be made observable without enabling personal recognition to be performed – e.g. in order for user interactions to be personalised. As a consequence, we may apply initially unlinkable pseudonyms (e.g. hashed IP addresses) to maintain relationships in user-adaptive systems. Beyond complete unidentifiability, unconditional anonymity as well implies complete untraceability (q.v. [7]). However, the question of linkability, being dependent on the use of PII, remains undecided[29] (see Table 2). Since users do not have to trust anyone so as to avoid accountability, the scope of trust contains no entities.

---

[28] In default of observable pseudonyms, linkability is, though conditionally, still ensured (due to identifiability).

**Remarks.** One may also achieve unconditional anonymity by introducing non-persistent (still initially unlinkable), transaction pseudonyms. Being different from transaction to transaction, this type, in contrast to the persistent one, does not support linkability. Moreover, as pointed out by Köhntopp and Pfitzmann [11], transaction pseudonyms (if used exactly once) can generate the same degree of anonymity as if there were no pseudonym at all.

## Summary

Table 1 and Table 2 comprehensively summarise all implications of what has been said in this section. As a remark: by defining a property as 'void', we indicate that it is just the inverse which is satisfied (e.g. void recognisability implies unrecognisability).

| LEVEL | | TYPE OF ANONYMITY | SCOPE OF TRUST | | RECOGNISABILITY | |
|---|---|---|---|---|---|---|
| DEG. | ABBR. | | LOWER BOUND | MAX. EXT. | TYPE | SOURCE |
| 0 | VA | void | system scope | – | unconditional | identifiability or traceability |
| 1 | AA | apparent | trusted set | system scope | unconditional | identifiability or traceability |
| 2 | RA | revocable | observee | trusted set | conditional | identifiability or traceability |
| 3 | FA | forfeitable | void | observee | conditional | identifiability |
| 4 | UA | unconditional | – | void | void | none |

*Table 1 – Class-level characterisation of anonymity*

| LEVEL | | TYPE OF LINKABILITY | TYPE OF LEGAL ACCOUNTABILITY | |
|---|---|---|---|---|
| DEG. | ABBR. | | | |
| 0 | VA | unconditional | direct | LOW LEVELS |
| 1 | AA | unconditional | direct or indirect | |
| 2 | RA | unconditional or conditional | indirect | HIGH LEVELS |
| 3 | FA | unconditional or conditional | direct or indirect | |
| 4 | UA | [undecided] | void | |

*Table 2 – Class-level characterisation of anonymity (cont.)*

---

[29] may either be conditional, unconditional, or void

Given that the scope of trust plays a central role in our approach, Figure 2 might prove useful for promoting better understanding of how the scope boundaries, in particular, evolve from level to level. With reference to services ensuring given types of anonymity, columns demarcated by dashed lines respectively illustrate (with relatively proportionate heights) the amount of trust that users must place in participants.

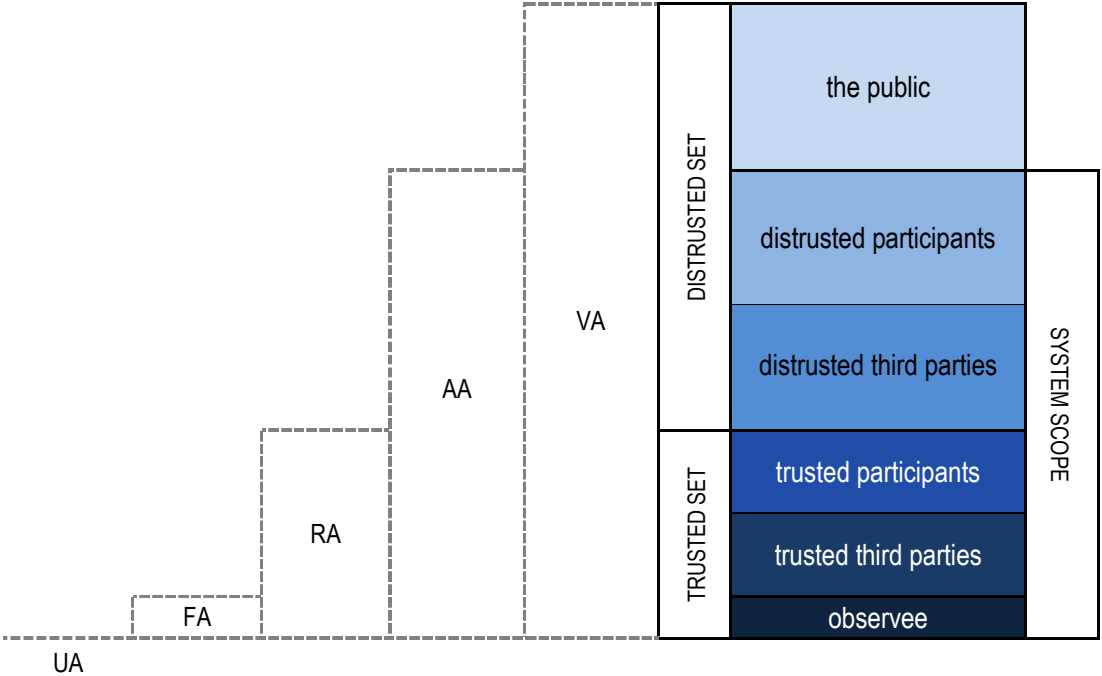

*Figure 2 – Illustration of the scope of trust*

## 5. CONCLUSION

Having studied earlier approaches, each of which views anonymity from a different aspect, we can safely say that either one is as relevant of its kind as any other. Nevertheless, as the scope of trust can certainly be of reasonable concern to the user, we do believe that our categorisation model may become, at least, a viable alternative (or complement) to extant schemes of classification. To get a bigger picture of the connections between all aforementioned taxonomies, let us take a look at Table 3. Since, by their very nature, these concepts can seldom be uniquely mapped to each other, we cannot generally establish one-to-one correspondences between them. In consequence, most mappings shown in Table 3 are rather approximate. It must also be underlined that, as a main advantage, our approach not only combines the benefits of the others, but as well remedies their deficiencies. Most notably, allowing for a more sophisticated comparison between anonymity services, it introduces the concept of forfeitable anonymity.

**Remarks.** As for Köhntopp and Pfitzmann's taxonomy, for simplicity, we confine the examination only to publicly observable pseudonyms. Otherwise, most rows should be crammed with additional types, which could, in turn, needlessly complicate understanding of relationships. On the other hand, parentheses indicate that the given level of anonymity can eliminate the need of using the respective pseudonym, however it does not exclude it practically.

| PRESENT | FLINN AND MAURER | KÖHNTOPP AND PFITZMANN | SEYS ET AL. |
|---|---|---|---|
| void anonymity | usual identification | public pseudonym | no anonymity or semi-anonymity |
| apparent anonymity | latent identification | initially non-public pseudonyms | |
| linkable revocable anonymity | | | conditional persistent anonymity |
| unlinkable revocable anonymity | | initially non-public transaction pseudonyms | conditional one-time anonymity |
| linkable forfeitable anonymity | ✗ | [30]initially unlinkable pseudonyms | ✗ |
| unlinkable forfeitable anonymity | ✗ | (initially unlinkable transaction pseudonyms) | ✗ |
| linkable unconditional anonymity | pen-name or anonymous identification | initially unlinkable pseudonyms | unconditional persistent anonymity |
| unlinkable unconditional anonymity | no identification | (initially unlinkable transaction pseudonyms) | unconditional one-time anonymity |

*Table 3 – Correspondences between taxonomies*

---

[30] The use of initially unlinkable pseudonyms, per se, is not sufficient to ensure (linkable) forfeitable anonymity.

# BIBLIOGRAPHY

[1] Douglas R. Stinson, "Cryptography: Theory and Practice", Third Edition, Chapman and Hall/CRC, 2005.

[2] A. Menezes, P. van Oorschot, and S. Vanstone, "Handbook of Applied Cryptography", CRC Press, 1996.

[3] D. Chaum and E. van Heyst, "Group signatures", Advances in Cryptology — EUROCRYPT '91, LNCS, vol. 547, pp. 257–265, Springer-Verlag, 1991.

[4] M. O. Rabin, "Digitalized signatures", Foundations of Secure Computation, pp. 155–168, Academic Press, 1978.

[5] S. Goldwasser, S. Micali, and C. Rackoff, "The knowledge complexity of interactive proof-systems", 17th Annual ACM Symposium on Theory of Computing, pp. 291–304, ACM Press, 1985.

[6] L. Law, S. Sabett, and J. Solinas, "How to make a mint: the cryptography of anonymous electronic cash", National Security Agency Office of Information Security Research and Technology (Cryptology Division), 1996.

[7] D. Chaum, "Security without identification: transaction systems to make big brother obsolete", Communications of the ACM, vol. 28, no. 10, pp. 1030–1044, 1985.

[8] D. Chaum, "Showing credentials without identification: Signatures transferred between unconditionally unlinkable pseudonyms", Advances in Cryptology — EUROCRYPT '85, LNCS, vol. 219, pp. 241–244, Springer-Verlag, 1986.

[9] K. Ohta, T. Okamoto, and K. Koyama, "Membership Authentication for Hierarchical Multigroups Using the Extended Fiat–Shamir Scheme", Advances in Cryptology — EUROCRYPT '90, LNCS, vol. 473, pp. 446–457, Springer-Verlag, 1990.

[10] J. Guajardo, B. Mennink, and B. Schoenmakers, "Anonymous Credential Schemes with Encrypted Attributes", Cryptology And Network Security, LNCS, vol. 6467, pp. 314–333, Springer Science, 2010.

[11] A. Pfitzmann and M. Köhntopp, "Anonymity, Unobservability, and Pseudonymity — A Proposal for Terminology", Designing Privacy Enhancing Technologies 2000, LNCS, vol. 2009, pp. 1–9, Springer-Verlag, 2001.

[12] C. C. Aggarwal and P. S. Yu, "Privacy-Preserving Data Mining: Models and Algorithms", Advances in Database Systems, vol. 34, 2008.

[13] A. Evfimievski and T. Grandison, "Privacy-Preserving Data Mining", Handbook of Research on Innovations in Database Technologies and Applications: Current and Future Trends, pp. 527–536, Springer Science, 2009.

[14] C. Diaz, S. Seys, J. Claessens, and B. Preneel, "Towards measuring anonymity", Privacy Enhancing Technologies 2002, LNCS, vol. 2482, pp. 54–68, Springer-Verlag, 2003.

[15] A. Serjantov and G. Danezis, "Towards an Information Theoretic Metric for Anonymity", Privacy Enhancing Technologies 2002, LNCS, vol. 2482, pp. 41–53, Springer-Verlag, 2003.

[16] B. Flinn and H. Maurer, "Levels of Anonymity", Journal of Universal Computer Science, vol. 1, pp. 35–44, Springer-Verlag, 1995.

[17] A. M. Froomkin, "Anonymity and its Enmities", Journal of Online Law, art. 4, 1995.

[18] A. Kobsa and J. Schreck, "Privacy through pseudonymity in user-adaptive systems", ACM Transactions on Internet Technology, vol. 3, no. 2, pp. 149–183, 2003.

[19] S. Seys, C. Díaz, B. De Win, V. Naessens, C. Goemans, J. Claessens, W. Moreau, B. De Decker, J. Dumortier, and B. Preneel, "Anonymity and Privacy in Electronic Services – Requirement study of different applications", APES Project, Deliverable 2 , 2001.

[20] B. Gavish and J. H. Gerdes, "Anonymous mechanisms in group decision support systems communication", Decision Support Systems, vol. 23, iss. 4, 297–328, 1998.

[21] C. Díaz, V. Naessens, J. Claessens, S. Nikova, C. Goemans, M. Loncke, B. De Win, S. Seys, B. De Decker, J. Dumortier, and B. Preneel, " Anonymity and Privacy in Electronic Services – Applications requirements for controlled anonymity", APES Project, Deliverable 7, 2003.

[22] E. McCallister, T. Grance, and K. Scarfone, "Guide to Protecting the Confidentiality of Personally Identifiable Information (PII)", National Institute of Standards and Technology Special Publication 800-122, 2010.